\documentclass[aps,twocolumn,psfig,epsfig,floats,citesort]{revtex4}
\usepackage{dcolumn}
\usepackage{amssymb,amsmath}
\usepackage{epsfig}
\newcommand\euro{{\sffamily C%
    \makebox[0pt][l]{\kern-.70em\mbox{--}}%
    \makebox[0pt][l]{\kern-.68em\raisebox{.25ex}{--}}}~}

\def\be{\begin{equation}}
\def\ee{\end{equation}}

\def\begineq{\begin{equation}}
\def\endeq{\end{equation}}

\begin{document}
\bibliographystyle{prsty}

\begin{abstract}
According to Shoemaker, the
 ``impact of solid bodies is the most fundamental process that
has taken place on the terrestrial planets'' \cite{sho77}, as they
shape the surfaces of all solar system bodies. A lot of
information on this process has been extracted from remote
observations of impact craters on planetary surfaces. However, the
nature of the geophysical impact events is that they are
non-reproducible. Moreover, their scale is enormous and direct
observations are not possible. Therefore, we choose an alternate
and of course downscaled experimental approach in order to
guarantee reproducible results: We prepare very fine sand in a
{\it well defined} and {\it fully decompactified}
 state by letting gas bubble through it.
After turning off the gas stream, we let a steel ball fall on the
sand. The series of events in the experiments and corresponding
discrete particle simulations is as follows: On impact of the
ball, sand is blown away in all directions (``splash'') and an
impact crater forms. When this cavity collapses, a {\it granular
jet} \cite{tho01,mik02b} emerges and is driven straight into the
air. A second jet goes downwards into the  air bubble entrained
during the process, thus pushing surface material deep into the
ground. The air bubble rises slowly towards the surface, causing a
granular eruption. In addition to the experiments and the discrete
particle simulations we present a simple continuum theory to
account for the void collapse leading  to the formation of the
upward and downward jets. We show that the phenomenon is robust
and even works for {\it oblique} impacts: the upward jet is then
shooting {\it backwards}, in the direction where the projectile
came from.

\end{abstract}

\title{Impact}
\author{Detlef Lohse, Raymond Bergmann, Ren\'e Mikkelsen, Christiaan Zeilstra,
Devaraj van der Meer, Michel Versluis, Ko van der Weele, Martin
van der Hoef, and Hans Kuipers}
\address{
Faculty of Science  and J.\ M.\ Burgers Centre for
Fluid Dynamics, University of Twente, 7500 AE Enschede, The Netherlands\\
}

\date{\today}

\maketitle

\noindent


\begin{figure}[htb]
\begin{center}

\vspace*{-0.6cm}

\epsfxsize=.9\hsize \epsffile{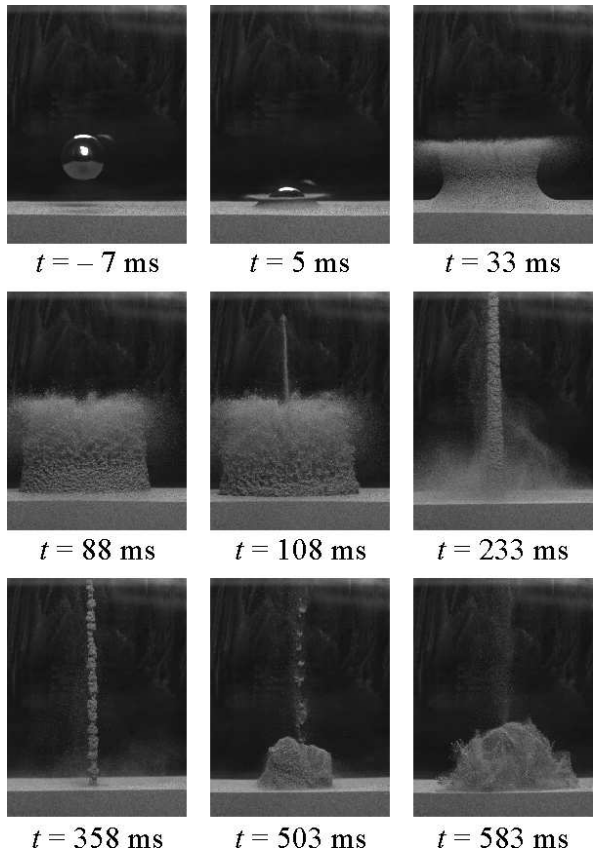}

\vspace{1mm}

\caption[]{Jet formation after the impact ($v_0=2.43m/s$)
of a steel ball
of $R_0 = 1.25cm$ on loose very
fine sand.
The jet in this experiment  exceeds
 the release height of the ball.
Frames 2-4: splash; frames 5-6: a jet emerges; frame 7: clustering
within the jet; frames 8-9: granular eruption at the surface. }
\label{fig_3d_exp}
\end{center}
\end{figure}


\begin{figure}[htb]
\begin{center}

\vspace*{-0.6cm}

\epsfxsize=.9\hsize \epsffile{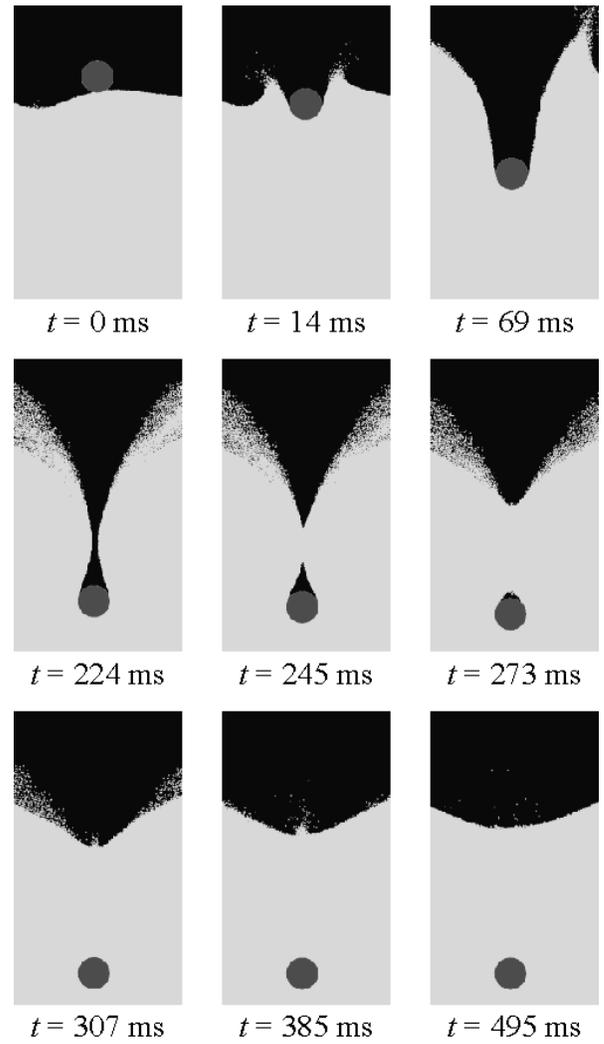}

\vspace{1mm}

\caption[]{Cut through the quasi-2D
discrete particle simulation.
Frames 1-3: the impact of the disk on the particles; Frames 4-6:
the collapse of the void; Frames 7-8: the upward jet (which is less
pronounced than in the 3D experiments).}
\label{fig_3d_sim}
\end{center}
\end{figure}

\begin{figure}
\begin{center}

\vspace*{-0.6cm}

\epsfxsize=.9\hsize \epsffile{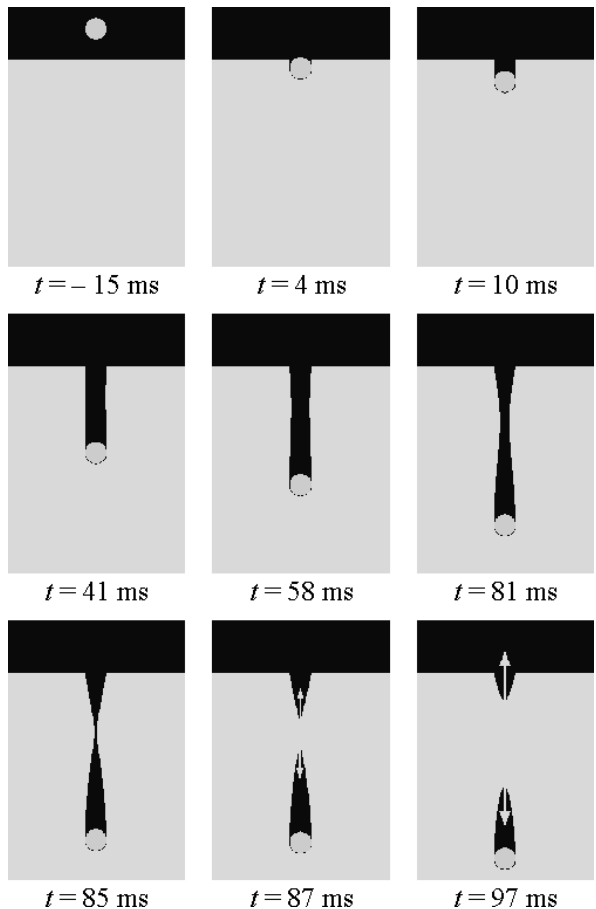}

\vspace{1mm}

\caption[]{Cross-section  of  the 3D-void collapse
following from our Rayleigh-type model, for the same
impact velocity and ball radius as in figure 1.
The void is pressed together by the ``hydrostatic'' pressure from
the side, leading to a singularity and an upward and downward jet.
}
\label{fig_3d_theory}
\end{center}
\end{figure}


\begin{figure}[htb]
\begin{center}
\epsfxsize=.7\hsize \mbox{\hspace*{.05 \hsize}
\epsffile{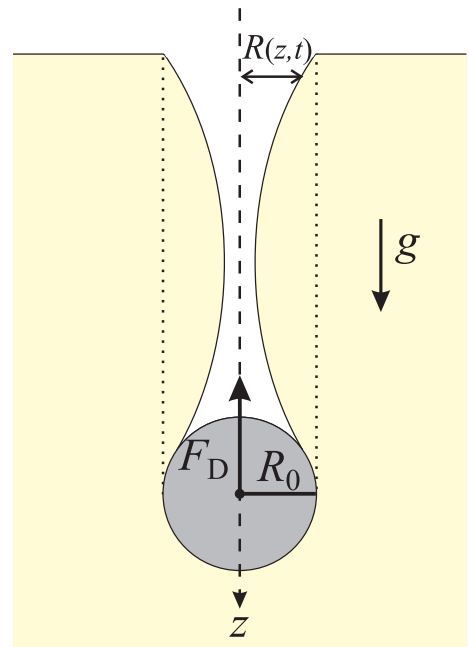} }

\caption[]{ Sketch of the void collapse. When the accelerated sand
grains from the sidewalls of the cylindrical cavity collide on the
axis of the cavity, two jets are formed: One downward into the
entrained air bubble formed above the sphere,
and one upward straight into the air. }
\label{fig_height}
\end{center}
\end{figure}

\begin{figure}[htb]
\begin{center}

\vspace*{-0.6cm}

\epsfxsize=.9\hsize \epsffile{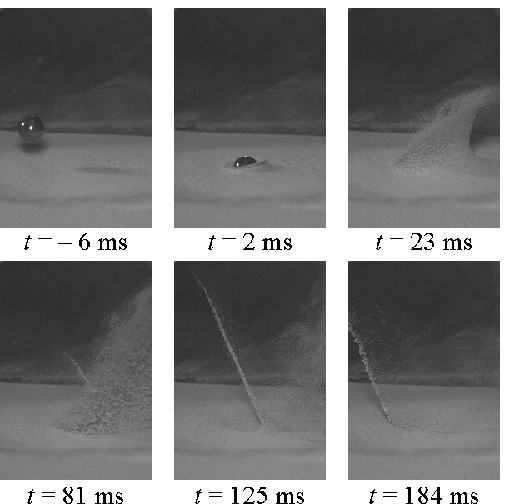}

\vspace{1mm}

\vspace{1mm} \caption[]{Impact of the steel ball on soft, loose
sand under an angle of approximately 45 degrees. Frames 2-4:
forwardly directed splash; frames 4-5: a backward jet emerges;
frame 6: clustering within the jet.} \label{fig_3d_45degrees}

\end{center}
\end{figure}

It has long been known that jets can be created when a ball or a
fluid droplet impacts on a fluid surface
\cite{wor08,ogu90,pro93b,hog98,loh03a}. S.\ Thoroddsen and A.\
Shen found  similar jets on impact of lead spheres on monodisperse
spherical glass beads \cite{tho01}.
We did similar experiments on
fine sand, but found it hard to achieve
quantitatively reproducible results, presumably due to the random
nature of the force-chain-networks in the
granular material \cite{liu95,jae96,kad99,her01}.
Therefore, in order to prepare a well-defined initial state,
we decompactify  and
 homogenize extremely  fine sand
(average grain-size of about $40\mu m$; grains are non-spherical)
 by blowing air through it
via a perforated bottom plate.
The height of the
sand bed above the bottom plate is typically 25-40cm. The air
is slowly turned off before the experiments and the grains are
left to settle in an extremely loose packing
with the force-chains either broken or substantially weakened.
We call this a ``fluid-like'' state. Impact events on this
well-prepared fine sand will be gravity-dominated.
We let a steel
ball (radius $R_0=1.25$cm) fall from various heights (up to $1.5$m)
onto the sand
and observe the dynamics of the sand with a digital high-speed
camera (up to 2000 frames per second).

The series of {\it visible} events is as follows (see figure
\ref{fig_3d_exp}): First, the ball vanishes in the sand and a
crown-like {\it splash} is created. Inhomogeneities develop in the
crown, due to the inelastic particle-particle
interaction (figure 1, frames 3-5). Then, after a while, a {\it
jet}
 shoots out of the sand at the position
of impact. In all our experiments the jet height exceeds the
release height of the ball. (The jets of ref.\ \cite{tho01} never
reach the release height,  because the sand is less fine
and much less  decompactified.) While the upper part of the jet is
still going upwards, in the lower parts the
inelastic particle-particle collisions lead to
density inhomogeneities in the jet (figure 1, frames
7-8).  These inhomogeneities resemble those of the surface tension
driven Rayleigh-instability of a water jet, even though there is no
surface tension in granular matter. Finally, after about half a
second, a {\it granular eruption}
 is seen at the position of impact, resembling
a volcano (figure 1, frames 8-9). The collapsing jet first leaves
a {\it central peak} in the crater{\footnote{Similar peaks
are observed in many
craters of the terrestrial planets \cite{sch92b,wal03}}},
but the
granular eruption violently erases this peak.


How does the jet form? To find out what is going on {\it below the
surface} of the sand, we (i) performed direct numerical
simulations, (ii) redid the experiments in two dimensions, meaning
that we replaced the ball by a cylinder (with axis parallel to the
surface and orthogonal to the side plates)
which we let fall into a bed of sand between two
transparent plates, and (iii) employed the analogy to jet formation
in fluids \cite{wor08,ogu90,pro93b,gla96,hog98,loh03a}.

(i) In the discrete particle simulations, the sand particles are
modeled as spheres which interact via inelastic ``soft-sphere''
collision rules. The interaction of the particles  with the
surrounding air is included via empirical drag force relations
\cite{li03}. Since the maximum number of particles that can be
simulated is presently of the order of one million, we can perform
only quasi-two-dimensional simulations, where the thickness of the
sand bed between the parallel plates is  eight
grains{\footnote{The 3D simulations we did are too strongly
affected by finite size effects. Nevertheless, also for these
simulations a jet emerges.}}. Altogether, the calculation includes
N= 1.3 million homogeneous beads of density $1000 kg/m^3$ and
diameter $500\mu m$  (i.e., approximately a factor 10 larger than
in experiment) in a container of 24cm $\times$ 0.4cm ground area
and a sand bed height of about 17cm. The  beads are pre-fluidized
with air, just as in the experiments, and then a 1.5cm diameter
ball of density $3500 kg/m^3$ is dropped onto the  beads with an
impact velocity of 2 m/s. The series of events can be seen in
figure \ref{fig_3d_sim}, revealing the jet formation process
invisible in figure \ref{fig_3d_exp}: The impacting ball creates a
void which is then pressed together through the ``hydrostatic''
pressure from the side. \footnote{In our experiments we are in the
region where the pressure increases linearly with depth
\cite{dur99}.} At small depth the ball passes early, meaning an
early start of the void collapse, which however is weak due to the
small ``hydrostatic'' pressure. Conversely, at larger depth the
collapse of the void begins later, but is stronger due to the
larger ``hydrostatic'' pressure. Somewhere in the middle the
collapse is finished first, and the void walls hit each other. It
is this singularity which leads to the formation of {\it two}
jets: One upwards and one downwards into an air bubble which was
entrained in the sand by the void collapse. The falling jet often
leaves a central peak in the crater (which in our 3D experiments
with the fine, decompactified sand is subsequently erased again by
the granular eruption). Note that the jet in the discrete particle
 simulations is much less pronounced than in
experiment. First, because the
 beads in the simulations are much larger than the sand
grains in the experiment, i.e. the sand bed is
less fluid-like and allows for less fine structure. Second,
the singularity due to
the focussing along the axis of symmetry is weaker in 2D and quasi-2D
experiments or  simulations than in 3D,
and the jet takes the form of a sheet.

(ii)
We performed such 2D jet formation experiments, by letting a
cylinder fall into decompactified sand between two transparent
plates, and observing the jet formation process from
the side
(see supplementary material and
ref.\ \cite{mik02b}). These experiments confirm
the above sketched series of
events. Again, the jet is   less pronounced than in  the 3D
experiments.
The entrained air bubble slowly rises in these experiments,
finally leading to a granular eruption at the surface, just as observed
in 3D.

(iii) The same series of events is also found after an
analogous impact of a steel ball or a falling disk on water
\cite{wor08,ogu90,pro93b,gla96,hog98,gau98,loh03a}.
We will employ this analogy below  in order to set
up a theoretical model.

Before we do so, we discuss the role of the ambient air for the
jet formation.
We redid the discrete particle simulations
with an air pressure reduced to nearly zero (vacuum), giving nearly
indistinguishable results for the jet formation.
The ambient air, however, can play a role during the evolution of
the jet, provided that the impact velocity $v_0$ is very high.
For (3D) experiments with very high impact velocities
we observed that
after the splash the crown goes
{\it inwards} rather than outwards, due to the pressure reduction behind
the fast projectile (Bernoulli's law). The crown in fact can fully
close and the jet then hits the closed crown, leading to an
explosion-like collision (see the supplementary material)
which spreads material all over the place.

To work out the essentials of the void collapse,
we now construct a ``minimal'' continuum mechanical model.
First,   the delay curve $z(t)$ of the ball in the
sand can be  obtained from a  simple force balance model involving
drag, gravity,  and added mass.  It describes the experimental
results obtained for  a falling ball equipped with a
thin tail rod, which allows
for easy depth measurements \cite{ber04}.
The delay curve $z(t)$ of the ball is inverted to
obtain
$t_{pass} (z)$, the time when the ball passes the
layer of sand at depth $z$. This sets the initial conditions for
the collapse of the two-dimensional void, namely $R(z,t_{pass})=R_0$
and $\dot R(z,t_{pass})=0$. Here, $R(z,t)$ is the time and depth
dependent radius of the void, see figure \ref{fig_height}.

Next, the collapse of the void formed by the ball has to be described.
It
is driven by the (``hydrostatic'') sand
pressure $p(z)$ at depth $z$. For small $z$ the pressure simply is
$p(z) =\rho_s g z$, for larger $z$ it saturates \cite{dur99}.
Here, $\rho_s$ is the sand density, assumed to be constant. If we
neglect the dissipative processes both between the different
layers of sand and between the sand grains in one layer, the
dynamics for fixed depth $z$ is determined by the Euler equation,
\be \rho_s (\partial_t v (r,t) + v(r,t) \partial_r v(r,t)) =
-\partial_r p(r,t). \label{euler} \ee Here,
$v(r,t)$ is the velocity
field in the sand. With continuity $\partial_r (r v(r,t))=0$,  and
with the boundary conditions $v(R(t),t) = \dot R(t)$ at the void's
wall and $v (R_\infty ,t) =0$ far away from the void, one obtains
a Rayleigh-type \cite{ray17,bre95}
ordinary differential equation
 for each $R(z,t)$, namely \be (R\ddot R + \dot
R^2) \log {R\over R_\infty } + {1\over 2} \dot R^2 = {1\over
\rho_s} p(z) = gz. \label{rayleigh} \ee
The radius $R_\infty $ is of the
order of the system size, but the results only weakly
(logarithmically) depend on this parameter. The dynamics following
from this Rayleigh-type model is shown in figure
\ref{fig_3d_theory}, resembling the void collapse in the
discrete particle
simulations figure \ref{fig_3d_sim}, in the  2D
experiments (see the supplementary material),
in experimental work on the void collapse in
transparent fluids \cite{wor08,ogu90,pro93b,gla96,hog98,loh03a},
in boundary integral simulations of the complete hydrodynamical
equations \cite{gau98}, and therefore presumably also in the 3D
experiments in sand shown in figure \ref{fig_3d_exp}.
Just before and at the singularity ($R(t)=0$ and diverging velocity),
the dynamics is determined by $R\ddot R + \dot
R^2 = 0$, which has the solution $R(t) \sim (t_s -t)^{1/2}$, where
$t_s$ is the time of the singularity. The velocity therefore
has a square-root divergence $\dot R(t) \sim (t_s -t)^{-1/2}$.

Having shown that the void collapse is driven by ``hydrostatic''
pressure, we now can  deduce scaling arguments \cite{ogu95},
for the limiting case of large impact velocity $v_0$, which
is the relevant one in the geophysical context.
The time up to void collapse in depth $z$ is the sum of the time
$z/v_0$ it takes the ball to get there and the collapse time itself,
which scales as $\sim R_0 / \sqrt{gz}$. The depth
$z_c$ where the walls of the void first touch (i.e., the position
of the singularity)
can be obtained from minimizing this sum with respect to $z$, resulting
in $z_c/R_0 \sim Fr^{1/3}$, where $Fr = v_0^2/(gR_0)$ is the Froude number.
From this one obtains that the time of the collapse $t_c$ scales as
$t_c \sim (R_0/v_0) Fr^{1/3} \sim \sqrt{R_0/g} Fr^{-1/6}$ \cite{ogu95}.
For large $v_0$ these
scaling laws are consistent both with our continuum model and with
our discrete particle simulations.

We now come to the question how things change under an {\it
oblique} impact {\footnote{In a geophysical context,  exactly
vertical impacts are of course very unlikely. 50\% of all impacts
on terrestrial planets occur with an angle between $30^o$ and
$60^o$ \cite{pie00}.}}? We performed experiments under an angle of
$45^o$. All other experimental conditions are as in figure
\ref{fig_3d_exp}. The series of events can be seen in figure
\ref{fig_3d_45degrees}. Remarkably -- but consistent with our
theoretical model
 -- the jet now points backwards, along the void created
by the impacting ball. The backwards jet is also observed
in the discrete particle simulations (see the supplementary material).
Note that this is different for an oblique impact on {\it water},
where the jet still goes upwards.

We conclude the paper with speculations on possible
implications of
our findings on the impact mechanism within the geophysical
context
\cite{rod78,kee77,mel89,sch92b,hol93,pie00}.
However, we would like to caution the reader because -- as
pointed out above -- lack of reproducibility of the details
typify geophysical events. Moreover, though our decompactifying procedure
minimized the {\it relative}
energy stored in the ground as compared to the energy
of the projectile, the {\it absolute} energy scales in our experiments
are of course very different as compared to geophysical events.
Nonetheless, we believe that the following speculations may stimulate
discussions in a geophysical context:
(i)  After the impact of a solid body on a planet,
it may be the upward jet and {\it not}
the splash which is the dominant source of planetary material
transferred into space \cite{kee77}.
Similarly, an oblique jet
resulting from an oblique impact allows for an enhanced sidewards
transport of material, as compared to the splash.
(ii) The
collapsing jet may contribute
to the central peak often found in impact craters \cite{sch92b,wal03}.
(iii) The downward jet will considerably change the layering of the
sediments  underneath a crater, as it provides a mechanism
how surface material can be transported deep into the ground (see
the supplementary material).
In addition, a granular eruption will rearrange the sediment.
Our suggested mechanism may shed new light on the sediment layering
data found underneath the Chicxulub crater, which is a source
of major controversy \cite{kel97b,smi99,mar01,kel03}.

\vspace{0.5cm}
\noindent
{\it Acknowledgement:}
The work is part of the research  program of
FOM, which is financially supported by NWO, and R. B.,
R. M., C. Z., D. v. d. M., and M. V.
 acknowledge
financial support.


\end{document}